# SimAMC: A Fast and Accurate Simulator for Resistive Memory-Based Analog Matrix Computing with Non-Idealities

Junbin Long

Institute for Artificial Intelligence, and School of Integrated Circuits, Peking University, Beijing, China

Mu Zhou

Institute for Artificial Intelligence, and School of Integrated Circuits, Peking University, Beijing, China

Saitao Zhang

Institute for Artificial Intelligence, and School of Integrated Circuits, Peking University, Beijing, China

Yongxiang Li

Institute for Artificial Intelligence, and School of Integrated Circuits, Peking University, Beijing, China

Zhong Sun

Institute for Artificial Intelligence, and School of Integrated Circuits, Peking University, Beijing, China

Analog matrix computing (AMC) circuits leverage resistive memory arrays to perform matrix operations in a massively parallel manner, providing an efficient approach for accelerating data-intensive tasks. However, hardware non-idealities severely impact computational accuracy, making early-stage simulation vital for reliable performance estimation and design optimization. While open-loop circuits for matrix-vector multiplication are well-studied, closed-loop AMC circuits—which solve matrix equations—are computationally more complex and substantially more sensitive to non-idealities, complicating their simulation. In this work, we present SimAMC, a simulator for resistive memory-based closed-loop AMC circuits. SimAMC is capable of modeling matrix inversion and eigenvector solving in the presence of key non-idealities, including device programming error, data conversion error, thermal noise, operational amplifier input offset, and interconnect resistance. For real-valued matrix computing circuits, an alternating iterative algorithm is designed. SimAMC's effectiveness is validated through comparison with SPICE, showing excellent agreement while also demonstrating a speedup of several orders of magnitude.

CCS CONCEPTS: • **Hardware → Electronic design automation.**

Additional Keywords and Phrases: analog computing, simulator, RRAM, resistive memory, non-ideality

## 1 INTRODUCTION

Matrix computing forms the foundation of modern scientific computing and data-intensive applications. Matrix-vector multiplication (MVM) dominates deep-learning workloads, while matrix inversion (INV) supports structured linear-system solvers in applications such as robotic inverse kinematics, and eigenvector computation (EGV) enables principal component analysis, spectral analysis, and large-scale ranking methods such as PageRank. However, traditional digital computing is becoming increasingly inadequate for modern matrix workloads due to the high computational complexity of digital matrix operations, the physical limits of CMOS scaling [1], and the memory-wall bottleneck inherent to von Neumann architectures [2].

In recent years, resistive memory-based analog matrix computing (AMC) has emerged as a promising solution for modern matrix workloads, leveraging massive parallelism and in-memory processing to reduce computational complexity and mitigate the memory-wall bottleneck [3, 4]. By employing resistive memory cross-point arrays and operational amplifiers (OPAs) to configure local or global feedback loops, AMC circuits perform a variety of matrix computations. Open-loop AMC circuits without global feedback implement MVM [5, 6], whereas closed-loop AMC circuits with global feedback enable the solution of matrix equations, such as INV and EGV [7]. Both open-loop and closed-loop AMC circuits can perform matrix computations in a single step, achieving significantly low time complexity [8, 9].

In AMC circuits, non-idealities such as device programming errors, noise, OPA input offset, and interconnect resistance can severely degrade computational accuracy. Therefore, accurate early-stage simulations are essential for performance estimation and circuit optimization. Numerous studies have proposed simulators and architectures for MVM circuits operating under non-ideal conditions, some of which are becoming widely adopted standards [10-21]. For open-loop MVM circuits, the circuit topology is feedforward, and the input-output relationship is linear and explicit. Consequently, prior works have typically modeled the effects of noise and interconnect resistance by superimposing row-wise and column-wise contributions or by independently perturbing conductance and voltage values [10, 11]. Even when more accurate modeling approaches are employed, the absence of global feedback allows the node equations to be rearranged into an iterative form that can be efficiently solved using standard fixed-point iterative methods [17].

In contrast, closed-loop AMC circuits are inherently more complex because the presence of global feedback fundamentally alters both the circuit behavior and the associated simulation challenges [8, 9]. The output voltages are directly fed back into the array through OPAs, creating a strong bidirectional coupling between the array node voltages and the circuit outputs. When non-idealities such as noise and interconnect resistance are taken into account, this coupling prevents the use of simple superposition or fixed-point formulations. Instead, accurate modeling requires explicitly leveraging OPA properties (*e.g.*, virtual short conditions) and deriving a globally coupled system of node equations. This leads to a large, sparse, yet tightly coupled linear system whose solution must be obtained through careful mathematical formulation and efficient numerical solving, rather than through straightforward iterative approaches.

Simulation of closed-loop AMC circuits can be performed using SPICE (Simulation Program with Integrated Circuit Emphasis), which typically relies on modified nodal analysis (MNA) to solve large sparse linear systems [22]. However, this process is computationally expensive, and the complexity grows further when accounting for non-idealities such as thermal noise and interconnect resistance, making SPICE unsuitable for large-scale simulations. Directly incorporating these effects into the SPICE netlist leads to a sharp increase in simulation time with matrix size, and in some cases renders simulation infeasible. Although a few studies have recently proposed simulators for closed-loop AMC circuits [23, 24], they do not account for the impact of non-idealities on circuit behavior. Therefore, there is a pressing need for a simulator that can efficiently and accurately evaluate how non-idealities affect the accuracy of closed-loop AMC circuits.

In this work, we introduce SimAMC, a simulator for resistive memory-based closed-loop AMC circuits. To capture the effects of non-idealities, we develop non-idealities model for both INV and EGV circuits. Device programming errors, data conversion errors, thermal noise, OPA input offset are represented using random variables and incorporated into the mathematical framework, while explicit formulas are derived to account for interconnect resistances. As a result, the outputs of SimAMC inherently reflect the combined impact of these non-idealities. Building on models of positive-valued matrix computing circuits, we further propose a dual-array differential scheme and an alternating iterative algorithm to handle real-valued matrix computing circuits. To validate SimAMC, we evaluate the effect of each non-ideality individually as well as collectively, and compare the results against SPICE under identical conditions. The results show that SimAMC is in close agreement with SPICE, while offering a substantial runtime advantage.

## 2 OVERVIEW AND MAPPING METHODS

### 2.1 Overview of SimAMC

The overview and workflow of SimAMC are shown in Figure 1. It supports Monte Carlo simulations, enabling users to perform statistical analysis of the effects of device programming errors and noise. The parameter module provides a unified configuration of circuit types, input parameters, non-idealities and circuit-related settings. The data generator produces batches of input text files, each containing different input matrix. The mapping module converts input matrices into conductance matrices in the cross-point arrays and maps input vectors to voltages. During this process, the device model adjusts conductance values to reflect programming errors. Once mapping is complete, SimAMC executes the selected circuit using a mathematical model, which incorporates data conversion errors, thermal noise, and interconnect resistance. Simulation results are saved in multiple output files. Finally, the result analysis module computes the relative error between simulated and ideal outputs and generates plots of batch results and errors.

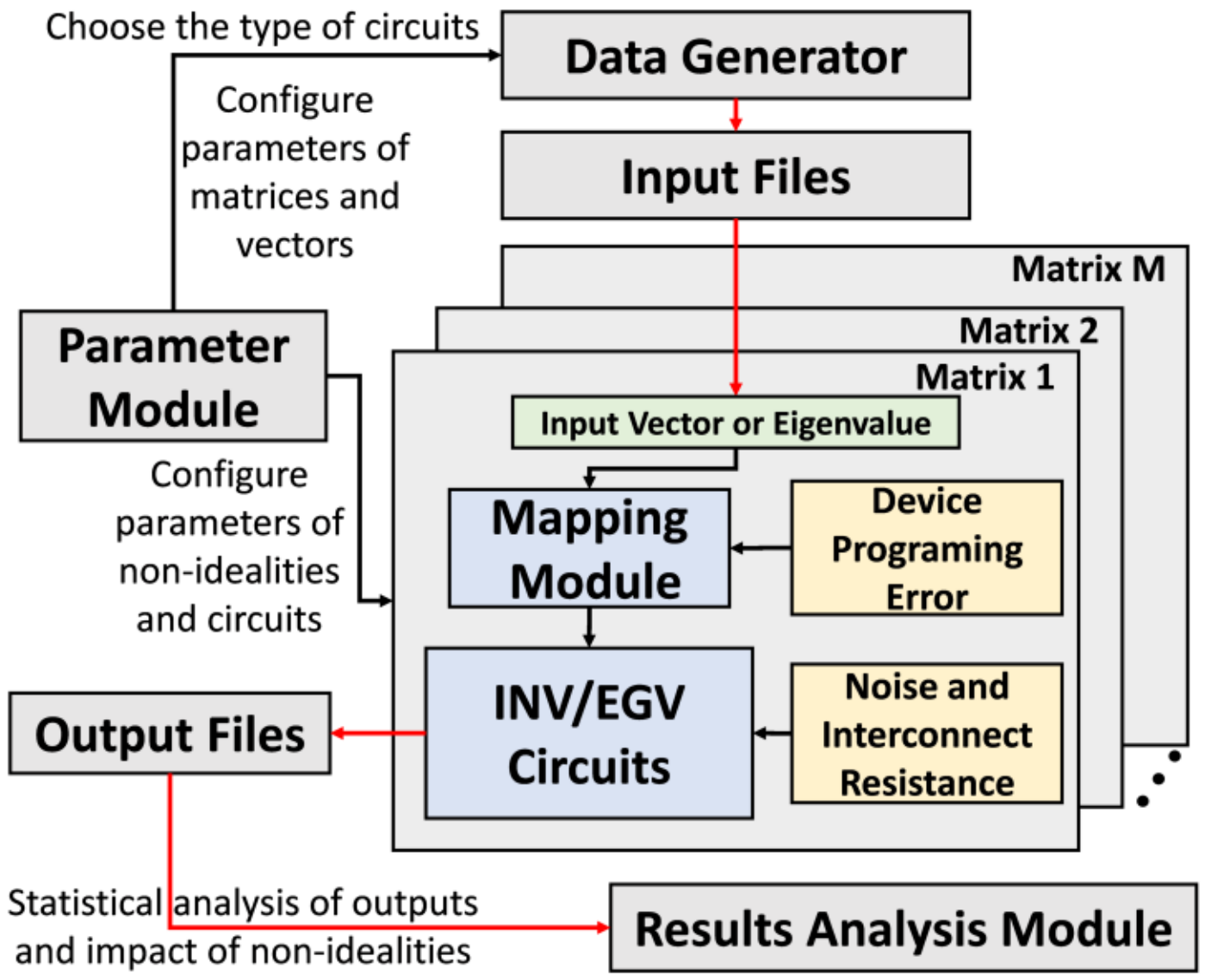

Figure 1: Overview and workflow of SimAMC.

## 2.2 Mapping Methods of SimAMC

Figure 2(a) shows the INV circuit [7] without non-idealities, which is used to solve systems of linear equations:

$$\boldsymbol{Ax} = \boldsymbol{b} \tag{1}$$

The $N \times N$ matrix $\boldsymbol{A}$ is mapped onto the conductance $\boldsymbol{G}$ of a cross-point array, and the vector $-\boldsymbol{b}$ is applied as the input voltage $\boldsymbol{v}_{in}$ . The circuit output voltage $\boldsymbol{v}_{out}$ corresponds to the solution of Eq. (1), *i.e.*, $\boldsymbol{x} = \boldsymbol{A}^{-1}\boldsymbol{b}$. The EGV circuit [7], shown in Figure 2(b), is used to solve eigenvectors of matrix $\boldsymbol{A}$, *e.g.*, the solution of equation

$$\boldsymbol{Ax} = \lambda \boldsymbol{x} \tag{2}$$

where $\lambda$ is a known eigenvalue of $\boldsymbol{A}$ and is mapped to the conductance of feedback resistors in the transimpedance amplifiers. To ensure the stability of the AMC circuits, the input matrices are chosen to be positive definite [9, 25]. The AMC circuits are always stable for positive definite matrices, and especially for well-conditioned ones, they can effectively tolerate programming errors and noise.

In matrix computing, elements can take arbitrary values, whereas physical quantities in circuits are restricted to specific ranges. To address these limitations, we apply the mapping procedures shown in Algorithm 1 and Algorithm 2 for the positive matrix INV and EGV circuits, respectively. The known quantities in the original matrix equations $\boldsymbol{A}'\boldsymbol{x}' = \boldsymbol{y}$ and $\boldsymbol{A}'\boldsymbol{x} = \lambda'\boldsymbol{x}$ can be specified over arbitrary ranges. In the INV circuit, a scaling factor $\alpha$ is used to limit the input voltage to a maximum of $\alpha$ V. The mapping from the scaled value to conductance is based on the unit conductance $G_0$. In addition, the output voltage of the EGV circuit is normalized using the Euclidean norm $\|\cdot\|_2$ to obtain a unit eigenvector.

Since conductance values can only be positive, we adopt a method that maps the real-valued matrix onto two distinct cross-point arrays using analog inverters. As shown in Figure 3, using the INV circuit as an example, the real-valued matrix $\boldsymbol{A}$ is split into a positive matrix $\boldsymbol{A}^+$ and a negative matrix $\boldsymbol{A}^-$, such that $\boldsymbol{A} = \boldsymbol{A}^+ - \boldsymbol{A}^-$. This scheme is also applied to the real-valued EGV circuit. To ensure proper splitting, a small bias $\delta$ is introduced. The splitting method is as follows:

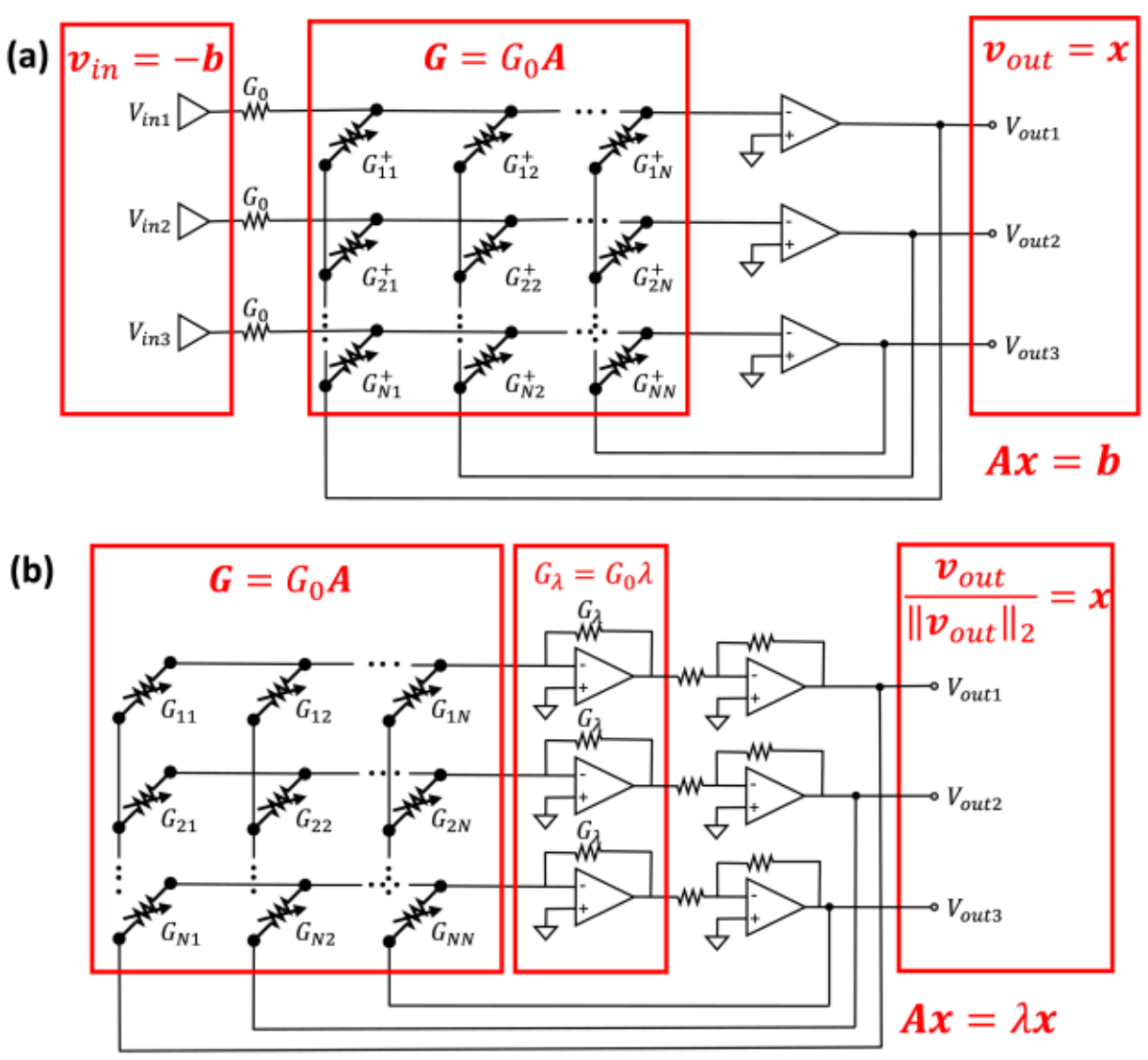

Figure 2: Schematic of (a) the INV circuit and (b) the EGV circuit without non-idealities.

$$
A_{ij}^{+} = \begin{cases} A_{ij} + \delta, A_{ij} \geq 0 \\ \delta, A_{ij} < 0 \end{cases}
$$
$$
A_{ij}^{-} = \begin{cases} \delta, A_{ij} \geq 0 \\ -A_{ij} + \delta, A_{ij} < 0 \end{cases} \quad (3)
$$

**Algorithm 1:** Mapping Process for the INV Circuit

**Input** : Original input matrix $\boldsymbol{A}'$, input vector $\boldsymbol{y}$.
**Output:** Solution $\boldsymbol{x}'$ of the equation $\boldsymbol{A}'\boldsymbol{x}' = \boldsymbol{y}$.

1 **Step 1:** Normalize: $\boldsymbol{A} = \dfrac{\boldsymbol{A}'}{\max(\boldsymbol{A}')}$, $\boldsymbol{b} = \dfrac{\alpha \boldsymbol{y}}{\max(\boldsymbol{y})}$;
2 **Step 2:** Map: $\boldsymbol{G} = G_0\boldsymbol{A}$, $\boldsymbol{v}_{\text{in}} = -\boldsymbol{b}$;
3 **Step 3:** Simulate: $\boldsymbol{G}\boldsymbol{v}_{\text{out}} = G_0(-\boldsymbol{v}_{\text{in}})$;
4 **Step 4:** Recover: $\boldsymbol{x} = \boldsymbol{v}_{\text{out}}, \boldsymbol{x}' = \dfrac{\max(\boldsymbol{y})}{\alpha \cdot \max(\boldsymbol{A}')}\boldsymbol{x}$;

**Algorithm 2:** Mapping Process for the EGV Circuit

**Input** : Original input matrix $\boldsymbol{A}'$, input eigenvalue $\lambda'$.
**Output:** Solution $\boldsymbol{x}$ of the equation $\boldsymbol{A}'\boldsymbol{x} = \lambda'\boldsymbol{x}$.

1 **Step 1:** Normalize: $\boldsymbol{A} = \dfrac{\boldsymbol{A}'}{\max(\boldsymbol{A}', \lambda')}$, $\lambda = \dfrac{\lambda'}{\max(\boldsymbol{A}', \lambda')}$;
2 **Step 2:** Map: $\boldsymbol{G} = G_0\boldsymbol{A}$, $G_\lambda = G_0\lambda$;
3 **Step 3:** Simulate: $\boldsymbol{G}\boldsymbol{v}_{\text{out}} = G_\lambda \boldsymbol{v}_{\text{out}}$;
4 **Step 4:** Recover: $\boldsymbol{x} = \dfrac{\boldsymbol{v}_{\text{out}}}{\|\boldsymbol{v}_{\text{out}}\|_2}$;

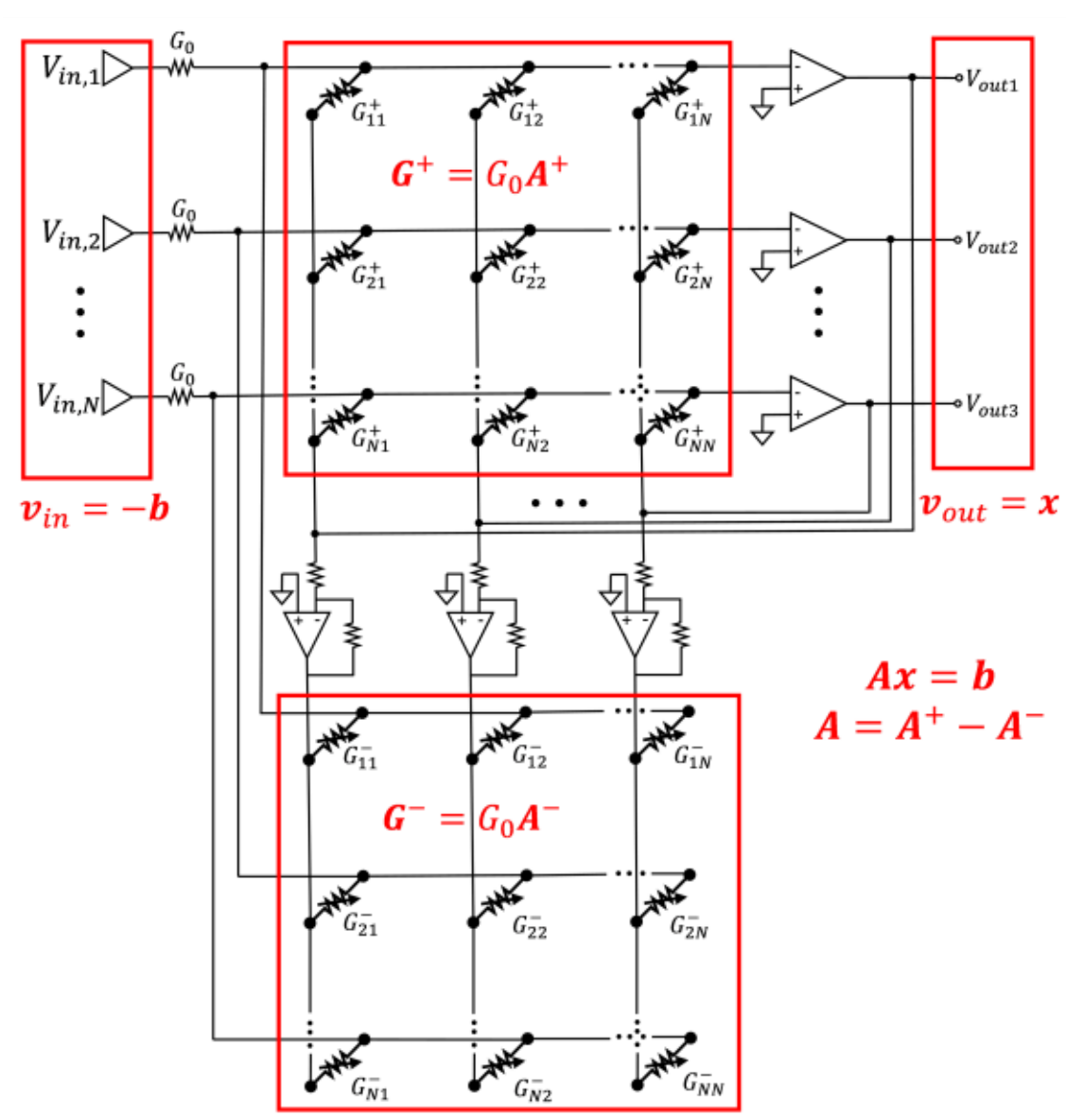

Figure 3: Schematic of the INV circuit with both positive and negative matrix elements.

where $A_{ij}$ denotes the element in the $i$-th row and $j$-th column of matrix $\boldsymbol{A}$, while $A_{ij}^{+}$ and $A_{ij}^{-}$ are defined analogously. In SimAMC, $\delta$ is typically set to satisfy the following equation:

$$\frac{A_{max} + \delta}{\delta} = \frac{G_{max}}{G_{min}} \tag{4}$$

where $G_{max}$ and $G_{min}$ denote the maximum and minimum conductance values of devices, respectively.

## 3 NON-IDEALITIES MODEL

To assess the impact of non-idealities in practical circuits on the accuracy of analog matrix computation, we develop a unified model that incorporates device programming error, data conversion error, thermal noise, OPA input offset, and interconnect resistance. The schematic of this model is shown in Figure 4. Directly modeling non-idealities with the conventional EGV circuit topology in Figure 2(b) makes it difficult to achieve convergence to a stable solution. To address this, we adopt an optimized EGV circuit in which one global feedback path is disconnected and a constant voltage $V_0$ is applied [26], as shown in Figure 4(b). To further extend the model for real-valued matrix computing, we employ an alternating iterative algorithm with positive and negative cross-point arrays. It should be noted that, unlike SPICE simulators that support dynamic circuit analysis and can provide detailed performance metrics such as transient behavior and power consumption, the proposed SimAMC framework is specifically designed for static simulation. The primary objective of this work is to investigate the impact of circuit non-idealities on the steady-state computational accuracy of AMC circuits. For this purpose, static simulation is sufficient to accurately capture and evaluate the accuracy degradation caused by various non-ideal factors, while maintaining a lightweight and computationally efficient simulation framework.

### 3.1 Overview of SimAMC

*3.1.1 Device Programming Error.* We model the devices based on measured resistive random-access memory (RRAM) data, but it is applicable to all resistive memories, which can be regarded as a two-terminal resistive device, including RRAM, phase change memory (PCM), magnetoresistive random-access memory (MRAM) and ferroelectric tunnel junction (FTJ) [27, 28]. During the RRAM programming process, deviations often arise between the target and actual conductance values. These errors stem from intrinsic device non-idealities, particularly the stochastic formation and rupture of conductive filaments [29], leading to device-to-device variations. Across a large number of devices, these variations typically follow a statistical distribution. Here, we model them using a Gaussian distribution, as illustrated in Figure 4.

During the mapping process, the target conductance values $G_{target,ij}$ are replaced by actual conductance values sampled from the distribution: $G_{ij} \sim N(G_{target,ij}, \sigma^2)$, where $G_{target,ij}$=$G_0 A_{ij}$. The standard deviation $\sigma$ represents the programming error and is assumed to be proportional to $G_{target,ij}$. To determine a reasonable range for the programming error, we experimentally characterize the statistical distribution of conductance values when RRAM devices are programmed to the same target conductance. The transmission electron microscopy (TEM) image and schematic of the RRAM device used in our measurements are shown in Figure 5(a) and Figure 5(b), respectively. In practical matrix-computing applications, a write-verify scheme is typically employed to reduce the programming error [30, 31]. As shown in Figure 5(c), for 5-bit RRAM programmed using the write-verify scheme, each state achieves a variation within 1%–5%. Therefore, in SimAMC, $\sigma$ is set to 3% of the target conductance to represent the typical programming-error distribution after applying the write-verify scheme. These sampled conductance values are then used for circuit simulation, thereby incorporating the effect of programming errors.

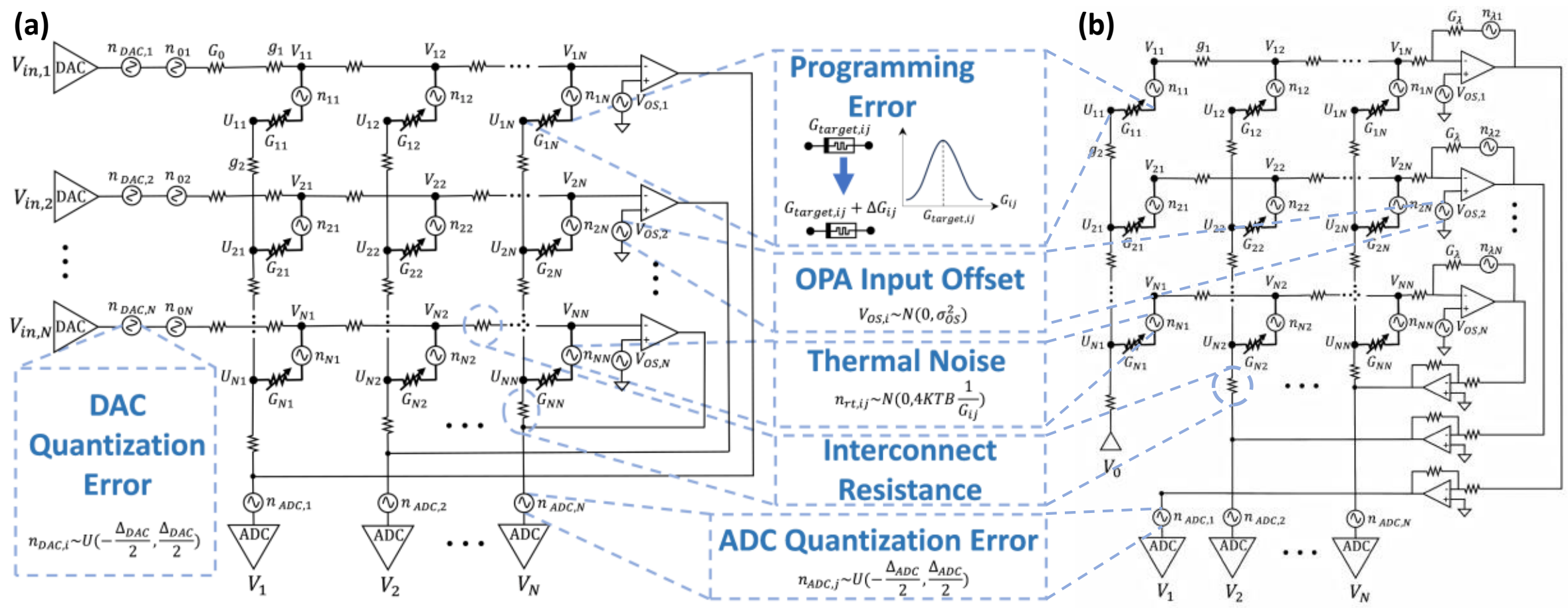

Figure 4: Schematic of the non-idealities model for (a) the INV circuit and (b) the EGV circuit. The figure depicts the device programming error, quantization error of DAC and ADC, device thermal noise, OPA input offset, and interconnect resistance.

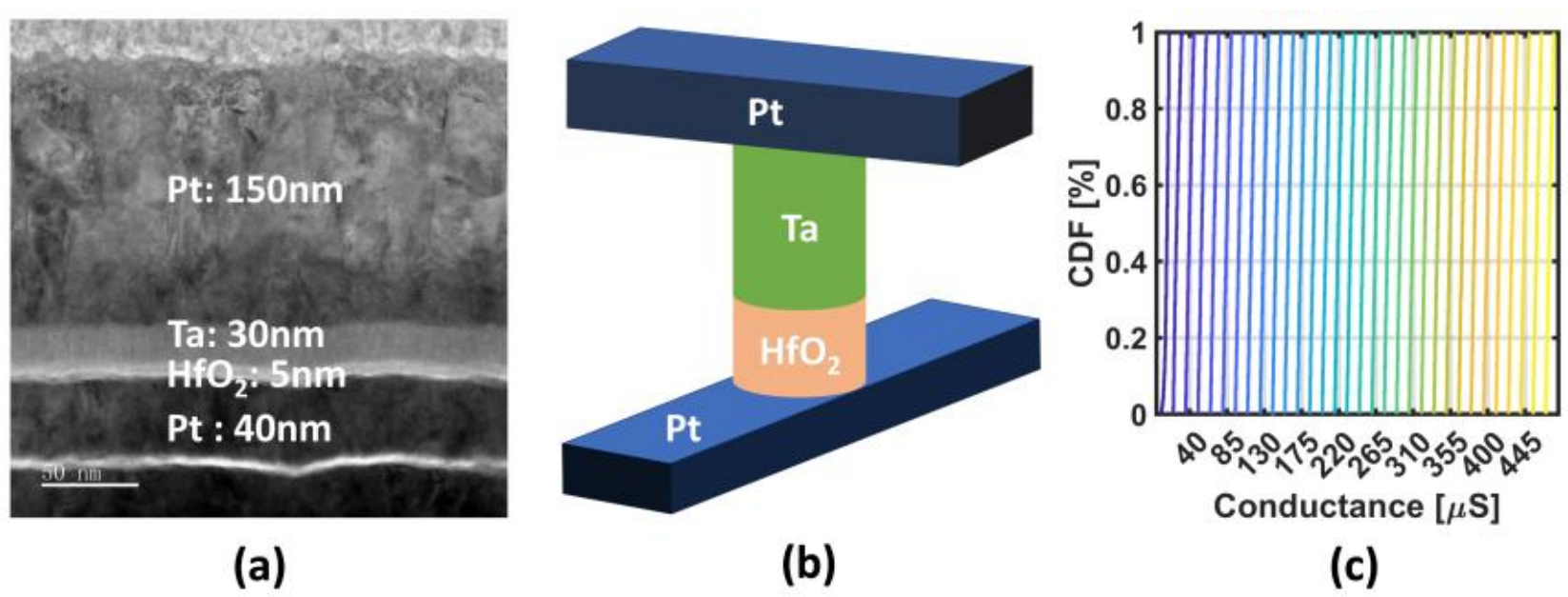

Figure 5: (a) The TEM image and (b) schematic of the RRAM used in our measurements. (c) RRAM conductance cumulative distribution function (CDF) when the device is closed-loop programmed using a write-verify scheme.

*3.1.2 Data Conversion Error.* SimAMC also models quantization errors in digital-to-analog converters (DACs) and analog-to-digital converters (ADCs) by introducing random voltage sources. The voltages, applied to the input of the $i$-th row and the output of the $j$-th column of the array, follow uniform distributions $n_{DAC,i} \sim U\left(-\frac{\Delta_{DAC}}{2}, \frac{\Delta_{DAC}}{2}\right)$ and $n_{ADC,j} \sim U\left(-\frac{\Delta_{ADC}}{2}, \frac{\Delta_{ADC}}{2}\right)$, respectively [32]. Here, $\Delta_{DAC}$ ($\Delta_{ADC}$) denotes the quantization step of DAC (ADC), defined as $\Delta_{DAC} = V_{DAC} \cdot 2^{-(B_{DAC}-1)}$ ($\Delta_{ADC} = V_{ADC} \cdot 2^{-(B_{ADC}-1)}$), where $V_{DAC}$ ($V_{ADC}$) is the full-scale output range and $B_{DAC}$ ($B_{ADC}$) is the resolution. Since the EGV circuit does not involve analog voltage input, DAC quantization error is not taken into account. As shown in Figure 4(a), DAC quantization error is added to the initial input voltage $V_{in,i}$, so that the actual input voltage of the $i$-th row in simulation is

$$V_{actual_in,i} = V_{in,i} + n_{DAC,i} \tag{5}$$

Similarly, the actual output voltage of the $j$-th column is

$$V_j = V_{ADC_in,j} + n_{ADC,j} \tag{6}$$

where $V_{ADC_in,j}$ is the ADC input. In this way, SimAMC's outputs reflect the impact of ADC quantization error.

*3.1.3 Thermal Noise.* Thermal noise arises from the thermal agitation of charge carriers and is related to device resistance. As the size of an array increases, the cumulative thermal noise also grows, exerting a stronger influence on AMC circuits.

Thermal noise is modeled as a random voltage source following a zero-mean Gaussian distribution [33]. This source is connected in series with the device and included in subsequent circuit simulations, as illustrated in Figure 4. If the device conductance $G_{ij}$ is known, the voltage across the cell, including the series-connected thermal noise source $n_{ij}$, can be expressed as:

$$\Delta V_{ij}{}' = \Delta V_{ij} + n_{ij} = \Delta V_{ij} + \sqrt{4KBT\frac{1}{G_{ij}}} \tag{7}$$

where $\Delta V_{ij}$ denotes the original voltage across the device, $K$ is the Boltzmann constant, $B$ is the equivalent noise bandwidth (ENBW), and $T$ is the absolute temperature. Here, the circuit bandwidth is approximated by the actual 3 dB bandwidth of the OPA, with the ENBW $B$ being 1.57 times the circuit bandwidth [34].

*3.1.4 OPA Input Offset.* The OPA input offset voltage primarily originates from device mismatches in the differential input pair and current mirror circuits caused by process variations, such as threshold voltage mismatch, mobility fluctuation, and channel dimension deviation. These mismatches introduce a small differential error at the input stage, which is subsequently amplified through the gain stages and ultimately appears as an equivalent DC voltage source at the input terminals. In addition, thermal effects and bias instability may also contribute to offset fluctuations. Since the dominant characteristic of the offset is its stochastic variation arising from fabrication non-idealities, the offset voltage is modeled as a Gaussian-distributed random variable: $V_{os,i} \sim N(0, \sigma_{os}^2)$. where $\sigma_{os}$ denotes the standard deviation of the offset voltage. Based on the schematic shown in Figure 4 and the virtual-short property of the OPA, the offset effect is incorporated into the circuit equations by setting

$$V_{iN} = V_{os,i} \tag{8}$$

thereby enabling the influence of the OPA offset to propagate through the subsequent circuit analysis.

*3.1.5 Interconnect Resistance.* The effect of interconnect resistance is incorporated by including it in the construction of the circuit node equations. We take the INV circuit as an example, with its circuit schematic shown in Figure 4(a). According to Kirchhoff's law, we derive the following equations for all $V_{ij}$ nodes:

$$\begin{gathered} G_{ij}\left(U_{ij} - V_{ij} - n_{ij}\right) = g_1\left(V_{ij} - V_{i,j+1}\right) + \frac{g_1 G_0}{g_1 + G_0}\left(V_{ij} - V_{in,1} - n_{DAC,i} - n_{0i}\right), j = 1 \\ G_{ij}\left(U_{ij} - V_{ij} - n_{ij}\right) = g_1\left(V_{ij} - V_{i,j+1}\right) + g_1\left(V_{ij} - V_{i,j-1}\right), j \neq 1 \text{ and } j \neq N \\ G_{ij}\left(U_{ij} - V_{ij} - n_{ij}\right) = g_1\left(V_{ij} - V_{i,j-1}\right), j = N \end{gathered} \tag{9}$$

where $g_1$ and $g_2$ represent the row and column conductance values of the interconnect resistances between adjacent nodes, respectively. $n_{0j}$ represents the thermal noise of the unit conductance $G_0$. Similarly, for all $U_{ij}$ nodes, we can derive

$$\begin{gathered} G_{ij}\left(V_{ij} + n_{ij} - U_{ij}\right) = g_2\left(U_{ij} - U_{i+1,j}\right), i = 1 \\ G_{ij}\left(V_{ij} + n_{ij} - U_{ij}\right) = g_2\left(U_{ij} - U_{i+1,j}\right) + g_2\left(U_{ij} - U_{i-1,j}\right), i \neq 1 \text{ and } i \neq N \\ G_{ij}\left(V_{ij} + n_{ij} - U_{ij}\right) = g_2\left(U_{ij} - V_j - n_{ADC,j}\right) + g_2\left(U_{ij} - U_{i-1,j}\right), i = N \end{gathered} \tag{10}$$

where $V_j$ corresponds to $\boldsymbol{v}_{out} = (V_1, V_2, \ldots, V_N)^{\mathrm{T}}$ in Algorithm 1. Eq. (9) involves $N^2$ unknowns, namely $V_{ij}$, while Eq. (10) contain $N(N+1)$ unknowns, consisting of $U_{ij}$ and $V_j$. By combining Eq. (9) with Eq. (10), all $U_{ij}$ can be eliminated, leading to the following matrix equation:

$$\boldsymbol{F}_{INV}(\boldsymbol{V}, \boldsymbol{V}_x) = \boldsymbol{0} \tag{11}$$

where $\boldsymbol{F}_{INV}$ is a function of the node voltage variables $V_{ij}$ and $V_j$, $\boldsymbol{V}$ is an $N \times N$ matrix composed of $V_{ij}$, and $\boldsymbol{V}_x = (0,0, \ldots, \boldsymbol{v}_{out})^{\mathrm{T}}$. According to the virtual short principle of the OPA, the input node voltage of OPA ideally satisfies $V_{iN} = 0$. Thus, $N$ unknowns can be eliminated, allowing $\boldsymbol{V}$ and $\boldsymbol{V}_x$ to be combined into a single matrix variable $\boldsymbol{F}_{INV}(\boldsymbol{\theta})$,

**Algorithm 3:** Solving Matrix Equations of INV Model with Positive Entries

**Input:** $\boldsymbol{v}_{in}, \boldsymbol{G}, G_0, g_1, g_2$
**Output:** $\boldsymbol{v}_{out}$
1 Initialize matrices $\boldsymbol{V}_{in} \leftarrow 0$, $\boldsymbol{\theta}_0 \leftarrow 0$, $\boldsymbol{V} \leftarrow 0$, $\boldsymbol{V}_x \leftarrow 0$;
2 $\boldsymbol{V}_{in}[:, 1] \leftarrow \boldsymbol{v}_{in}$;
3 Generate random noise matrices: $\boldsymbol{n}_0, \boldsymbol{n}$;
4 **solve_pos:**
5 Compute $\boldsymbol{F}_{INV}(\boldsymbol{\theta}_0)$ and $\boldsymbol{J}_{INV}(\boldsymbol{\theta}_0)$ with $\boldsymbol{\theta}_0$;
6 Solve $\boldsymbol{F}_{INV}(\boldsymbol{\theta}_0) + \boldsymbol{J}_{INV}(\boldsymbol{\theta}_0)(\boldsymbol{\theta} - \boldsymbol{\theta}_0) = \boldsymbol{0}$ for $\boldsymbol{\theta}$;
7 $\boldsymbol{V}_x \leftarrow (\boldsymbol{\theta}[:, N])^T$;
8 $\boldsymbol{v}_{out} \leftarrow \boldsymbol{V}_x[N, :]$;

with $N^2$ variables corresponding to the node equations in Eq. (11). $\boldsymbol{F}_{INV}(\boldsymbol{\theta})$ consists of linear terms proportional to the variable matrix $\boldsymbol{\theta}$ (the unknown matrix $\boldsymbol{\theta}$ multiplied by known matrices), along with constant terms [35]. Since $\boldsymbol{F}_{INV}(\boldsymbol{\theta})$ is linear, without any higher-order (quadratic or above) terms in $\boldsymbol{\theta}$, its Jacobian matrix $\boldsymbol{J}_{INV}(\boldsymbol{\theta})$ is constant, and the Taylor expansion of $\boldsymbol{F}_{INV}$ about the initial variable $\boldsymbol{\theta}_0$ contains only the constant and first-order terms. Substituting the Taylor expansion into Eq. (10) yields:

$$\boldsymbol{F}_{INV}(\boldsymbol{\theta}_0) + \boldsymbol{J}_{INV}(\boldsymbol{\theta}_0) \cdot (\boldsymbol{\theta} - \boldsymbol{\theta}_0) = \boldsymbol{0} \tag{12}$$

where $\boldsymbol{\theta}_0$ is initialized as a zero matrix. Because the Jacobian is independent of $\boldsymbol{\theta}$, it can be computed directly from the matrices constructed with the known quantities in Eq. (9) and Eq. (10). The solution $\boldsymbol{\theta}$ is obtained by solving Eq. (12), from which $\boldsymbol{v}_{out}$ can be extracted. This output inherently reflects the effects of device programming error, data conversion error, thermal noise, OPA input offset, and interconnect resistance. The overall procedure is summarized in Algorithm 3. A similar algorithm is applied to the EGV circuit shown in Figure 4(b), yielding a solution with non-idealities. The sparsity of the Jacobian matrix greatly accelerates the solving process of Eq. (12), enabling rapid simulation for closed-loop AMC circuits with interconnect resistance.

### 3.2 Real-Valued Matrix Computing Circuits

For real-valued matrix computing circuits, we introduce an alternating iterative algorithm to model circuit non-idealities, as in the positive matrix computing case, as shown in Algorithm 4. Taking the INV circuit in Figure 3 as an example, non-idealities are incorporated in a similar way, except that two arrays are considered. This requires two sets of interconnect resistances and two sets of thermal noise sources. After solving Eq. (12) for the positive array, the node voltages in its first column and the outputs $V_j$ are used as inputs to the negative array. As with the positive array, we derive and simplify the node equations to obtain a matrix equation involving unknowns for the negative array, namely $\boldsymbol{F}^-_{INV}(\boldsymbol{V}^-) = \boldsymbol{0}$, where $\boldsymbol{V}^-$is an $N \times N$ matrix composed of nodes $V^-_{i1}$ of negative array corresponding to $V_{ij}$ in Figure 3. The solution is then obtained by solving

$$\boldsymbol{F}^-_{INV}(\boldsymbol{V}^-_0) + \boldsymbol{J}^-_{INV}(\boldsymbol{V}^-_0) \cdot (\boldsymbol{V}^- - \boldsymbol{V}^-_0) = \boldsymbol{0} \tag{13}$$

where $\boldsymbol{J}^-_{INV}$ is the Jacobian matrix of $\boldsymbol{F}^-_{INV}$. The solving processes of Eq. (12) and Eq. (13) correspond to the solve_pos and solve_neg functions in Algorithm 4, respectively. The current $I_i$ flowing from the negative array into the $i$-th row of the positive array is calculated as $I_i = g_1(V^-_{i1} - V_{i1})$. These currents are then applied to the positive array, updating the equation for $j = 1$ in Eq. (9) to

**Algorithm 4:** Solving Matrix Equations of INV Model with Both Positive and Negative Entries

```
   Input: v_in, G+, G-, G_0, g_1, g_2, max_iter, ε
   Output: v_out
1  Initialize matrices V_in ← 0, θ_0 ← 0, V+ ← 0, V_0- ← 0, I+ ← 0, V_x ← 0;
2  V_in[:, 1] ← v_in;
3  for k ← 1 to max_iter do
4  |  Generate random noise matrices: n_0, n+, n-;
5  |  θ ← solve_pos(θ_0, N, V_in, I+, G+, G_0, n_0, n+, g_1, g_2);
6  |  V_x ← (θ[:, N])^T;
7  |  V+ ← θ[:, 1 : N − 1];
8  |  V- ← solve_neg(V_0-, N, V+, V_x, G-, n-, g_1, g_2);
9  |  I+ ← g_1 × (V-[:, 1] − V+[:, 1]);
10 |  if ||θ − θ_0||_F < ε and ||V- − V_0-||_F < ε then
11 |  |  Output "Converged at iteration k";
12 |  |  break;
13 |  if k = max_iter then
14 |  |  Output "Fail to converge";
15 |  θ_0 ← θ;
16 |  V_0- ← V-;
17 v_out ← V_x[N, :];
```

$$G_{ij}\left(U_{ij}-V_{ij}-n_{ij}\right)=g_1\left(V_{ij}-V_{i,j+1}\right)+\frac{g_1G_0}{g_1+G_0}\left(V_{ij}-V_{in,1}-n_{DAC,i}-n_{0i}\right)-I_i \tag{14}$$

Eq. (12) is then solved again to update $V_j$ in the positive array. This process continues, with the positive and negative arrays alternately solving Eq. (12) and Eq. (13) while updating $V_j$ and $I_i$ until convergence is achieved according to the threshold $\varepsilon$ defined in the parameter module, yielding the final output $\boldsymbol{v}_{out}$. Consequently, the matrix computing solution contains both positive and negative elements. A similar alternating iterative algorithm can be applied to the EGV circuit, where the positive and negative arrays are also computed separately. The voltages at the nodes connecting the two arrays are iteratively updated in an alternating manner until convergence, yielding the voltages corresponding to the eigenvector. Owing to the rapid solution of each array, the alternating iteration remains efficient, even for real-valued matrix computing circuits.

## 4 VALIDATION AND EVALUATION

To verify that SimAMC can accurately capture various non-idealities, we validate the simulator by considering each non-ideality individually as well as all non-idealities collectively in both the INV and EGV circuits. In addition, we evaluate the computational efficiency of the simulator.

### 4.1 Validation of SimAMC

*4.1.1 Device Programming Error.* To evaluate the impact of device programming errors, we compare the simulation results of SPICE and SimAMC using 30 different 64×64 input matrices. In SPICE, a Monte Carlo simulation is performed in which device resistances in the cross-point array are perturbed directly according to the specified statistical variation model, whereas in SimAMC, Gaussian-distributed random numbers are generated for the mapped conductance values.

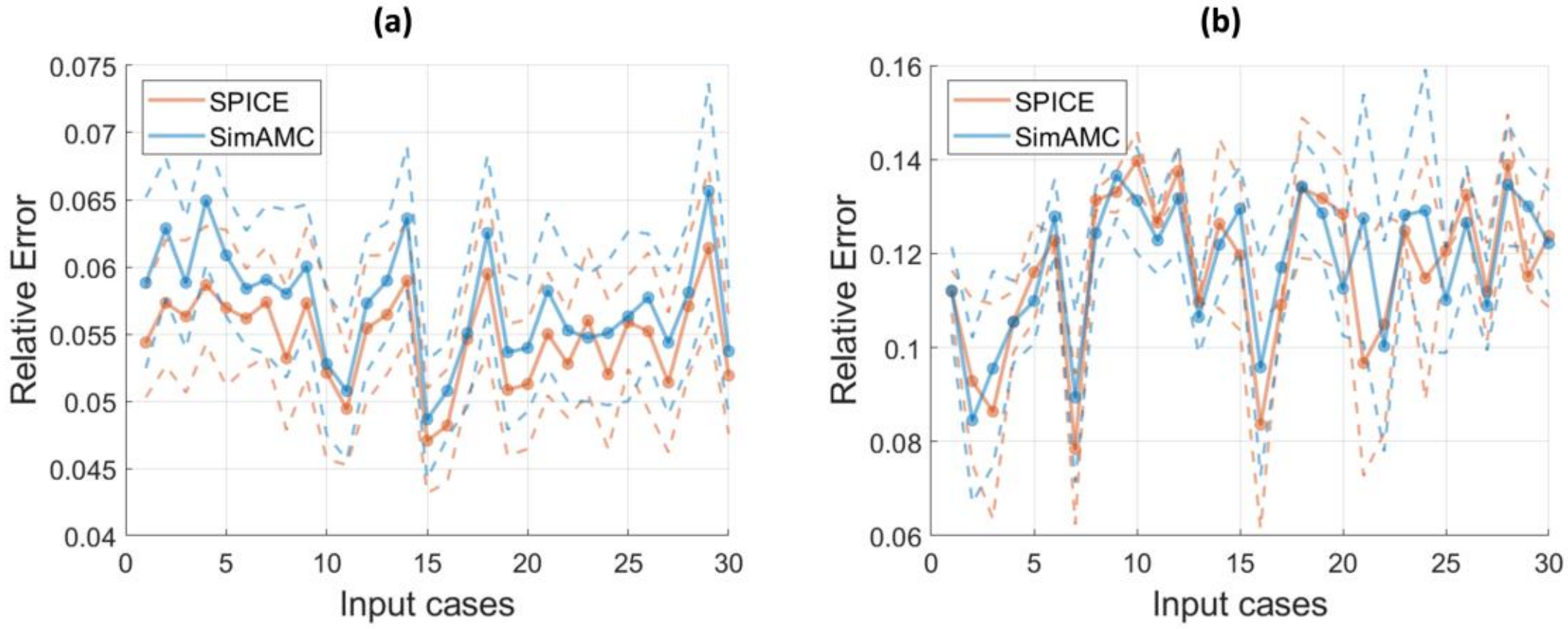

Figure 6: Relative error between simulation results and ideal matrix computing results for different matrices, taking device programming errors into account, obtained using device model and SPICE. The solid line represents the mean relative error from the batch simulations, while the dashed lines indicate the corresponding upper and lower bounds. (a) INV circuit. (b) EGV circuit.

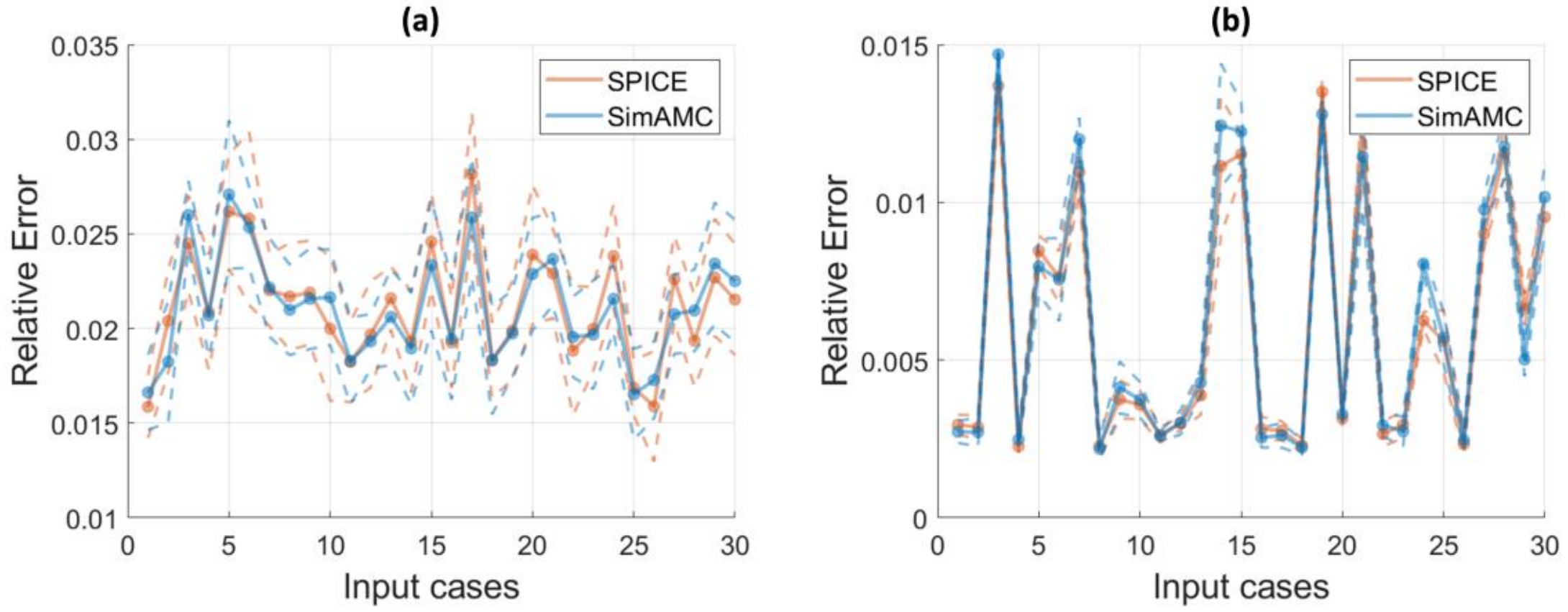

Figure 7: Relative error between simulation results and ideal matrix computing results for different matrices, taking data conversion error into account, obtained using DA/AD model and SPICE. (a) INV circuit. (b) EGV circuit.

The conductances range from 5 μS to 200 μS, with the standard deviation $\sigma$ set to 3% of the target conductance. Relative errors between the simulated and ideal results are calculated and averaged across the dataset, as shown in Figure 6. The errors from SimAMC closely match those from SPICE, with nearly identical trends across varying inputs, confirming that SimAMC effectively models device programming errors.

*4.1.2 Data Conversion Error, Thermal Noise, and OPA Input Offset.* In SPICE, DAC and ADC behavior is simulated using digital-to-analog and analog-to-digital conversion functions, while thermal noise and OPA input offset are modeled with random voltage sources. In SimAMC, random numbers are generated according to the specified distributions of data conversion error, thermal noise, and OPA input offset. For consistency, the standard deviation of the random voltage sources in SPICE is matched to that of the random numbers in SimAMC. The DAC and ADC are configured with a resolution of 12 bits. In the noise analysis, the ambient temperature is assumed to be 300K, while the equivalent noise bandwidth is approximated using the 3-dB bandwidth of the OPA. The OPA parameter configuration is based on the specifications of the AD823 [36]. Validation is performed via Monte Carlo simulations on 30 different 16×16 matrices, with each non-ideality tested individually (Figure 7, Figure 8 and Figure 9). The errors produced by our models are in close agreement with those from SPICE, exhibiting similar trends across varying inputs, which confirms that our models accurately reproduce the effects of data conversion error, thermal noise, and OPA input offset.

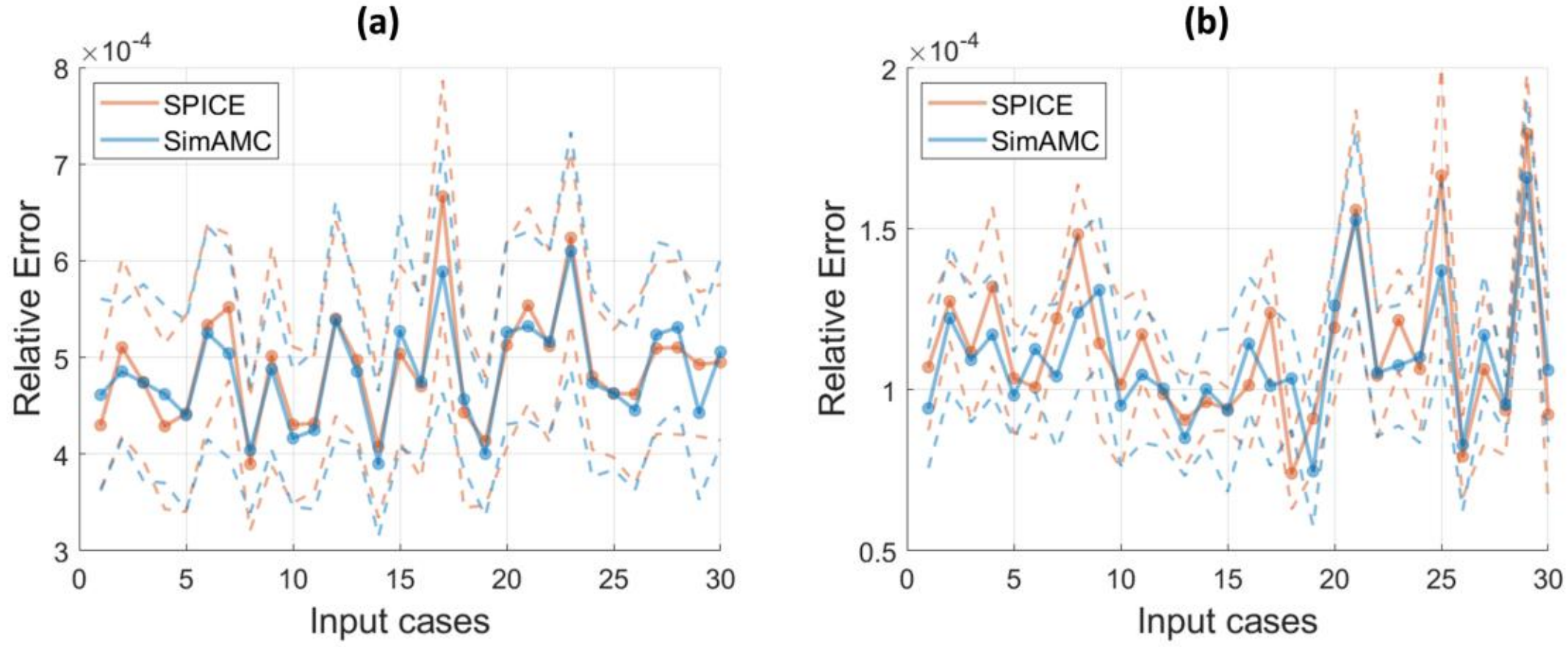

Figure 8: Relative error between simulation results and ideal matrix computing results for different matrices, taking thermal noise into account, obtained using noise model and SPICE. (a) INV circuit. (b) EGV circuit.

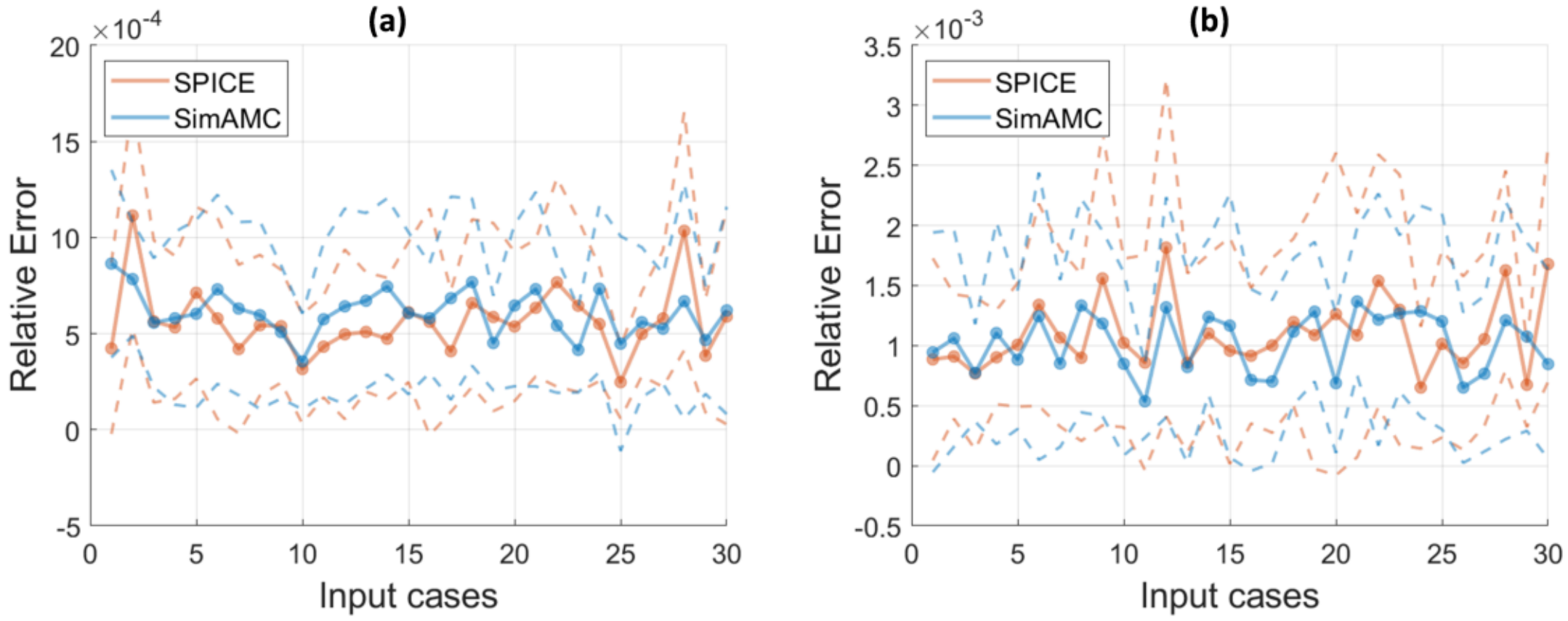

Figure 9: Relative error between simulation results and ideal matrix computing results for different matrices, taking OPA input offset into account, obtained using OPA input offset model and SPICE. (a) INV circuit. (b) EGV circuit.

*4.1.3 Interconnect Resistance.* SPICE simulates AMC circuits with interconnect resistance, while SimAMC calculates the output voltages using Eq. (11) with device programming error and noise disabled. Figure 10 compares the results under varying matrix sizes with an interconnect resistance of $5\,\Omega$. The outputs of SimAMC and SPICE show negligible differences, demonstrating that SimAMC accurately captures the effect of interconnect resistance.

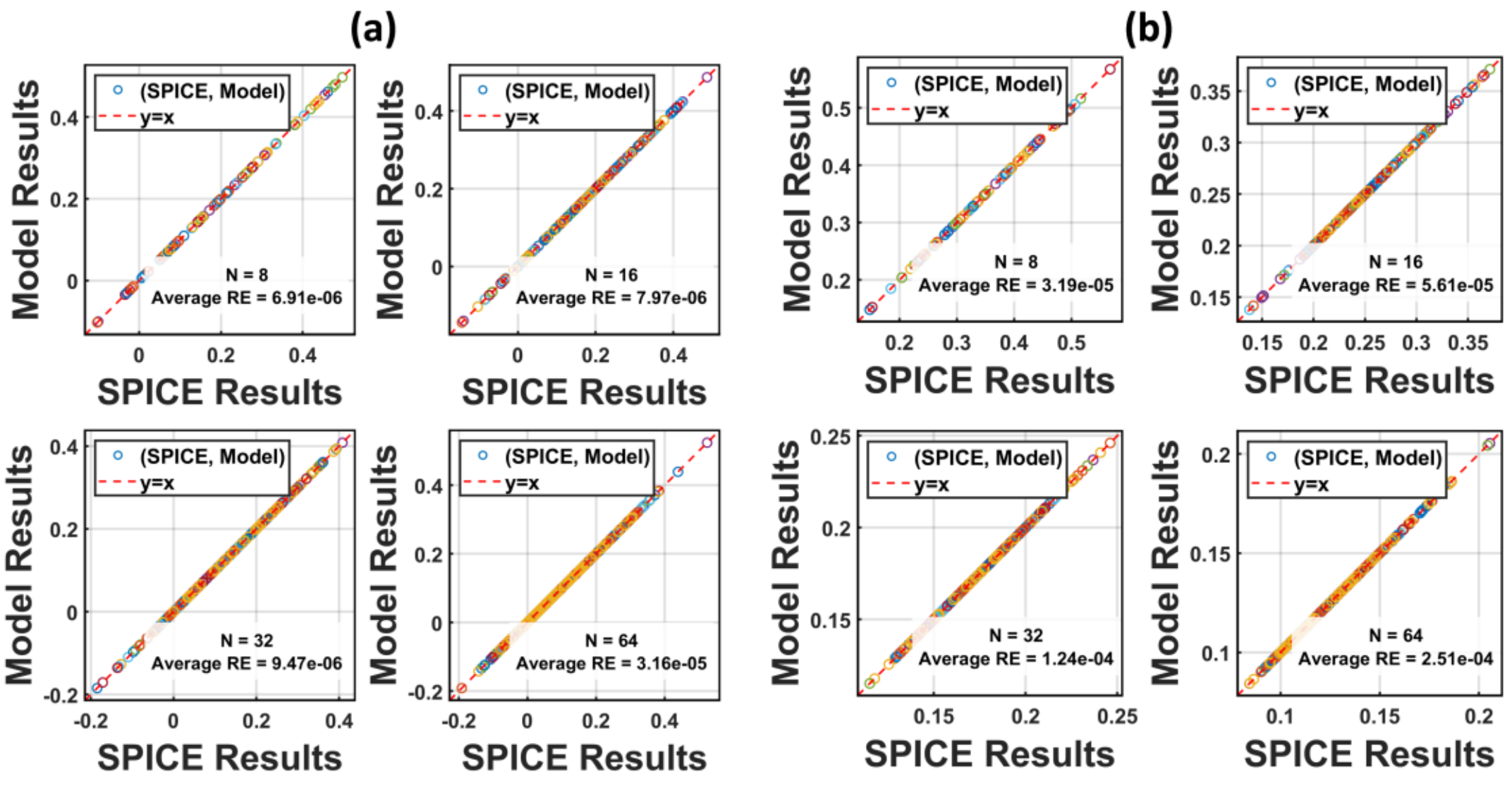

Figure 10: Comparison of simulation results using the inter-connect resistance model and SPICE. (a) INV circuit. (b) EGV circuit. The four subplots correspond to matrix sizes $N = 8, 16, 32, 64$. Each subplot contains ten sets of simulation results, marked with scatter points in different colors. The inset shows the average relative error (RE) for the ten sets of data in each subplot.

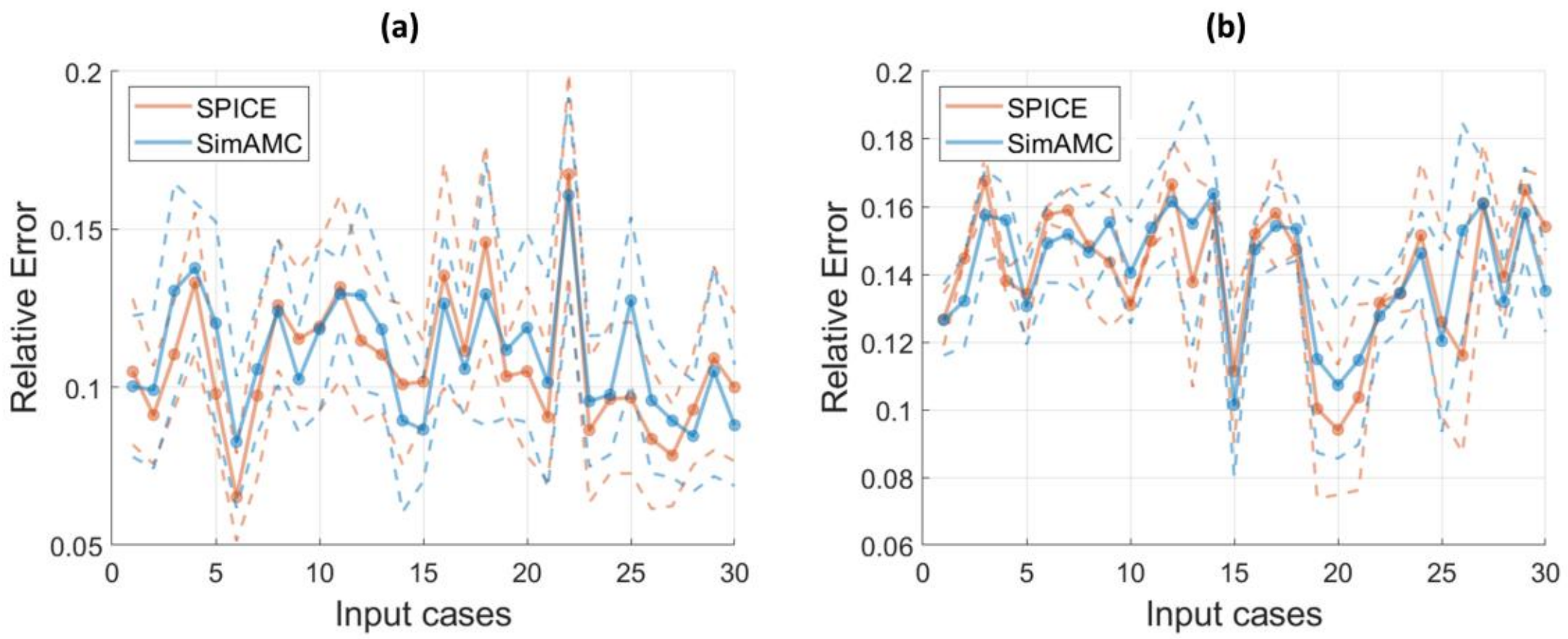

Figure 11. Relative error between simulation results and ideal matrix computing results for different matrices, taking all non-idealities into account, obtained using SimAMC and SPICE. (a) INV circuit. (b) EGV circuit.

*4.1.4 All Non-Idealities.* Finally, we validate SimAMC when all non-idealities are included simultaneously. SPICE performs Monte Carlo simulations on the netlist derived from the schematic in Figure 4, while SimAMC computes the outputs using Algorithm 3. As shown in Figure 11, the relative errors between the two are minimal, demonstrating excellent agreement.

## 4.2 Evaluation of SimAMC

To evaluate the runtime performance of SimAMC, we compare it with the LTSpice XVII simulator running on an Intel(R) Core(TM) i7-11800H CPU @ 2.30 GHz with 16 GB RAM. Under identical conditions, multiple simulations with different matrix sizes are performed, and the average runtime is calculated for both SimAMC and SPICE. For a fair comparison, ideal OPAs are employed in SPICE. Simulating non-idealities in SPICE is extremely time-consuming, primarily due to the large number of random noise sources in noise-inclusive simulations and the extensive resistors required in interconnect-resistance simulations. As a result, the maximum matrix size tested with SPICE is limited to 64×64. The simulation runtimes as a function of matrix size are shown in Figure 12, demonstrating that SimAMC achieves fast performance for AMC circuits with non-idealities, providing several orders of magnitude of acceleration over SPICE for both the INV and EGV circuits. Notably, although SimAMC requires iterative computation when simulating real-valued matrix computing circuits, it has also been demonstrated to provide a speedup over SPICE.

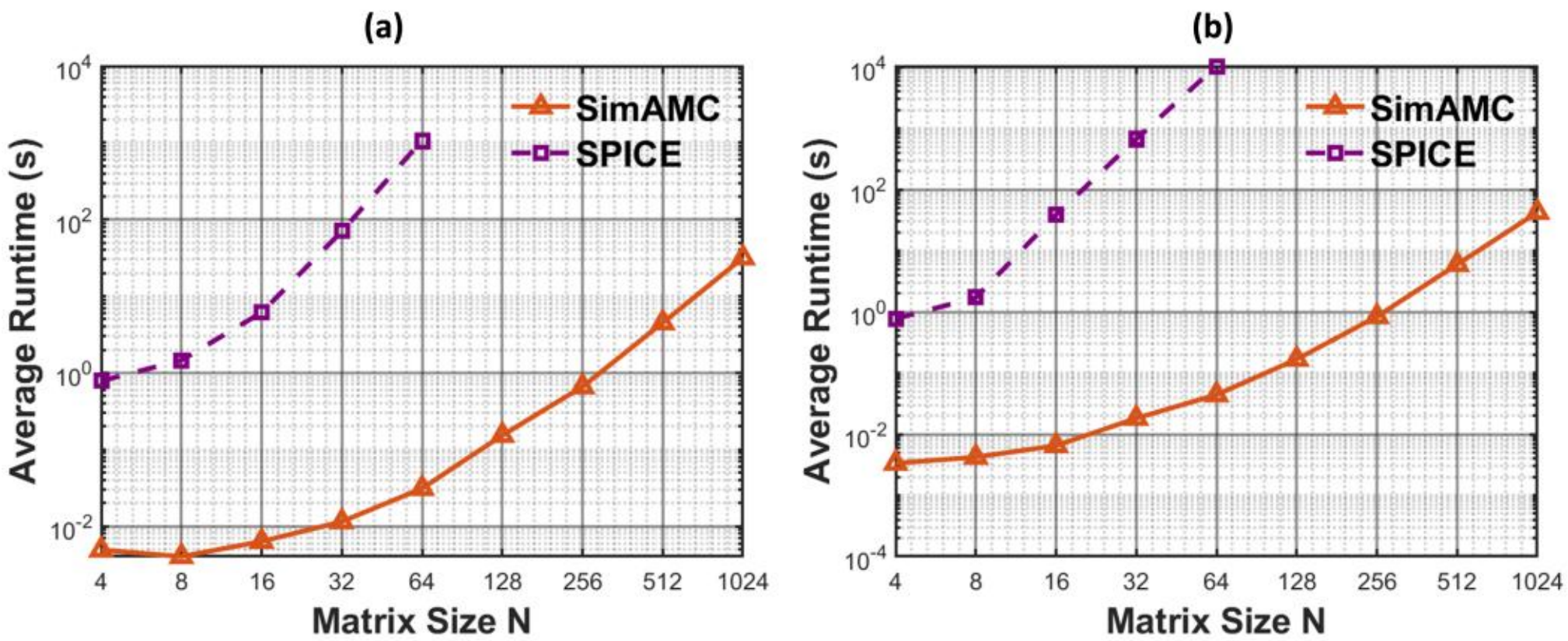

Figure 12. Runtime comparison between SimAMC and SPICE for (a) the INV circuit and (b) the EGV circuit, both including all non-idealities.

## 5 CONCLUSION

To conclude, we presented SimAMC, a resistive memory-based AMC circuits simulator, which can accurately simulate INV and EGV circuits with non-idealities, including device programming error, data conversion error, thermal noise, OPA input offset, and interconnect resistance. With the alternating iterative algorithm we designed, SimAMC can also be applied to real-valued matrix computing circuits. Under identical conditions, it achieves a simulation speedup of several orders of magnitude compared to SPICE. SimAMC provides circuit designers with an effective early-stage tool for evaluating and optimizing AMC circuits. Furthermore, this work represents a preliminary step toward the systematic development of AMC toolchains and simulators, offering valuable insights for comprehensive simulator design and practical AMC circuit applications.